\documentclass[%
 reprint,
 amsmath,amssymb,
 aps,
]{revtex4-1}

\usepackage{graphicx}
\usepackage{dcolumn}
\usepackage{bm}

\begin{document}

\preprint{APS/123-QED}

\title{Strong magnetization and Chern insulators in compressed graphene/CrI$_{3}$ van der Waals heterostructures}

\author{Jiayong Zhang$^{1,}$$^{3}$}\
\email{jyzhang@usts.edu.cn}
\author{Bao Zhao$^{2,}$$^{3}$}
\author{Tong Zhou$^3$}
\author{Yang Xue$^3$}
\author{Chunlan Ma$^1$}
\author{Zhongqin Yang$^{3,}$$^{4,}$}
\email{zyang@fudan.edu.cn}
\address{$^1$Jiangsu Key Laboratory of Micro and Nano Heat Fluid Flow Technology and Energy Application, $\Psi_{usts}$ Institute $\&$ School of Mathematics and Physics, Suzhou University of Science and Technology, Suzhou, Jiangsu 215009, China\\
$^2$College of Physical Science and Information Technology, Liaocheng University, Liaocheng 252000, China\\
$^3$State Key Laboratory of Surface Physics and Key
Laboratory for Computational Physical Sciences (MOE) $\&$ Department
of Physics, Fudan University, Shanghai 200433, China\\
$^4$Collaborative Innovation Center of Advanced Microstructures, Fudan University, Shanghai, 200433, China}

\date{\today}

\begin{abstract}
Graphene-based heterostructures are a promising material system for designing the topologically nontrivial Chern insulating devices. Recently, a two-dimensional (2D) monolayer ferromagnetic insulator CrI$_{3}$ was successfully synthesized in experiments [Huang \emph{et al.}, Nature 546, 270 (2017)]. Here, these two interesting materials are proposed to build a heterostructure (Gr/CrI$_{3}$). Our first-principles calculations show that the system forms a van der Waals (vdW) heterostructure, relatively facilely fabricated in experiments. A Chern insulating state is acquired in the Gr/CrI$_{3}$ heterostructure if the vdW gap is compressed to certain extents by applying an external pressure. Amazingly, very strong magnetization (about 150 meV) is found in graphene, induced by the substrate CrI$_{3}$, despite the vdW interactions between them. A low-energy effective model is employed to understand the mechanism. The work functions, contact types, and band alignments of the Gr/CrI$_{3}$ heterostructure system are also studied. Our work demonstrates that the Gr/CrI$_{3}$ heterostructure is a promising system to observe the quantum anomalous Hall effect at high temperatures (up to 45 K) in experiments.
\end{abstract}

\pacs{Valid PACS appear here}
\maketitle





\section{\textbf{INTRODUCTION}}
The Chern insulating state, exhibiting quantum anomalous Hall (QAH) effect, is a novel two-dimensional (2D) topological quantum state that is insulating in the bulk but host robust conducting edge states and exhibits quantized Hall conductivity in the absence of external magnetic field \cite{1,2,3,4}, which was first theoretically proposed by Haldane \cite{4}. The Chern insulating state has attracted considerable attention in condensed matter physics and materials science since the realization of the QAH effect in realistic material systems may greatly promote the development of low-power-consumption electronics devices. By now, the Chern insulating state has been theoretically predicted to occur in numerous material systems \cite{5,6,7,8,9,10,11,12,13,14,15,16}, such as mercury-based quantum walls \cite{5}, transition metal (TM) atoms doped topological insulator thin films \cite{6}, graphene and silicene based systems \cite{7,8,9,10,11,12,13}, 2D organic topological insulators \cite{14,15}, heavy atomic layers on magnetic insulators \cite{16}, and so on. However, the laboratory synthesis of these previously predicted QAH effect material systems is somewhat very difficult, making the experimental observations of the QAH effect still full of challenges. Till now, the QAH effect has been merely observed in TM atoms doped (Bi,Sb)$_{2}$Te$_{3}$ thin films \cite{17,18,19,20} at an extremely low temperature (such as 30 mK \cite{17}), seriously hindering the further development of this important research field. Therefore, predicting a QAH material system that can be facilely fabricated in experiments, with a large band gap and a high Curie temperature (\emph{T$_{c}$}), is of great significant.

Graphene is a unique 2D monolayer material with a honeycomb lattice formed by carbon atoms, which has become an ideal prototype material for engineering the QAH effect \cite{7,8} due to its special linear Dirac band dispersions and relatively mature technologies of sample growth and device fabrication. Theoretical studies showed that the Chern insulating state can be realized in graphene by introducing both magnetic exchange field and Rashba spin-orbit coupling (SOC) \cite{7}. Introducing a long-range ferromagnetism (FM) order in the target materials in experiments is one of the most important and tough tasks to carry out the Chern insulating state. Besides, the SOC of graphene must be enhanced, since the intrinsic SOC in pristine graphene is extremely weak \cite{21,22}. Previous studies indicated that the magnetic exchange field and the enhanced Rashba SOC can be acquired in graphene by depositing low-concentration TM atoms into graphene \cite{7,9}. The TM atoms tend to, however, gather into clusters \cite{23,24} on graphene surface, causing the experimental realization of the Chern insulating state in graphene based on this tactics hard. In graphene, the Chern insulating state was also predicted by depositing it onto a suitable magnetic insulator substrate \cite{12,13}, which may overcome the problem of the doping tactics. Hence, building graphene-based heterostructures, especially van der Waals (vdW) heterostructures \cite{25,26}, should be a very promising tactics to realize experimentally the Chern insulating state in graphene. With current experimental technologies, the vdW heterostructures containing different monolayer materials have been successfully fabricated in experiments, such as WSe$_{2}$/CrI$_{3}$ \cite{27}, graphene/MoS$_{2}$ \cite{28}, MoSe$_{2}$/WSe$_{2}$ \cite{29}, and so on.

In this work, we demonstrate that the compressed vdW heterostructure formed by a graphene and a monolayer CrI$_{3}$, denoted as Gr/CrI$_{3}$, is a Chern insulator. The bulk crystalline CrI$_{3}$ is a layered FM insulator with the adjacent interlayers joined by vdW interactions \cite{30}. The bulk CrI$_{3}$ has a relatively high Tc of 61 K with an out-of-plane easy axis \cite{30}. Recently, the 2D monolayer CrI$_{3}$ was synthesized in experiments with its FM and insulating behaviors preserved well, except for a slightly lower Tc of 45 K \cite{31}, compared to that of the bulk. By depositing graphene on the monolayer CrI$_{3}$, we find that the Dirac points of graphene layer are not located inside the bulk band gap of the 2D monolayer CrI$_{3}$ substrate, and the Chern insulating state cannot be realized with the equilibrium interface distance. Fortunately, if the distance between graphene and the monolayer CrI$_{3}$ is reduced properly by applying an external pressure perpendicular to the Gr/CrI$_{3}$ heterostructure plane, the Dirac points of graphene can be tuned flexibly into the bulk band gap of CrI$_{3}$, and the Chern insulator can be carried out. With the decrease of the interface distance, the Fermi level (E$_{F}$) of the Gr/CrI$_{3}$ heterostructure moves from the position near the bottom of the conduction bands to the top of the valence bands, indicating the variation from Ohmic contacts to Schottky contacts (n-tpye and p-type) at the Gr/CrI$_{3}$ interface. Also very interestingly, the substrate CrI$_{3}$ can induce a large magnetic exchange field (about 150 meV) in the top graphene layer, much stronger than that (about 70 meV) of the heterostructure of graphene on the (111) surface of BiFeO$_{3}$ with chemical bonding at the interface \cite{12}. The substrate can also enhance much the Rashba SOC in graphene, leading to the achieved QAH gap larger than 10 meV. Our findings demonstrate that the compressed vdW heterostructure of Gr/CrI$_{3}$ is an appropriate candidate system to experimentally realize the QAH effect at high temperatures (up to 45 K).

\section{\textbf{COMPUTATIONAL METHODS}}

The geometry and electronic structures of the vdW heterostructures Gr/CrI$_{3}$ are calculated by using the projector augmented wave formalism (PAW) \cite{32} based on density-functional theory (DFT), as implemented in the Vienna ab initio simulation package (VASP) \cite{33}. The exchange and correlation functional is described by using the Perdew-Burke-Ernzerhof generalized-gradient approximation (GGA-PBE) \cite{34}. To take into account the correlation effects of Cr 3\emph{d} electrons, the GGA+U method \cite{35} is adopted and the value of the on-site Coulomb interaction \emph{U} and exchange interaction \emph{J} are set to be 3.0 and 0.9 eV, respectively. The plane-wave cutoff energy is set to be 500 eV and a vacuum space of larger than 15 {\AA} is adopted to avoid the interaction between two adjacent heterostructure slabs. The convergence criterion for the total energy is set to be 10$^-6$ eV, and the Monkhorst-Pack k-point grids of $6 \times 6 \times 1$ are adopted. All atoms in the unit cell are allowed to move until the Hellmann-Feynman force on each atom is smaller than 0.01 eV/{\AA}. The vdW interaction functional with the method of Grimme (DFT-D2) \cite{36} is employed in the vdW heterostructure calculations. For the monolayer CrI$_{3}$, the hybrid functional HSE06 \cite{37} is employed to obtain its optimized geometry structures and ground-state electronic structures. The band structure of the vdW heterostructure with the equilibrium interface distance is also tested with the hybrid functional HSE06.

\section{\textbf{RESULTS AND DISCUSSION}}

\subsection{\textbf{Geometry and electronic structures}}

\begin{figure*}
\resizebox{12cm}{!}{\includegraphics*[62,30][518,393]{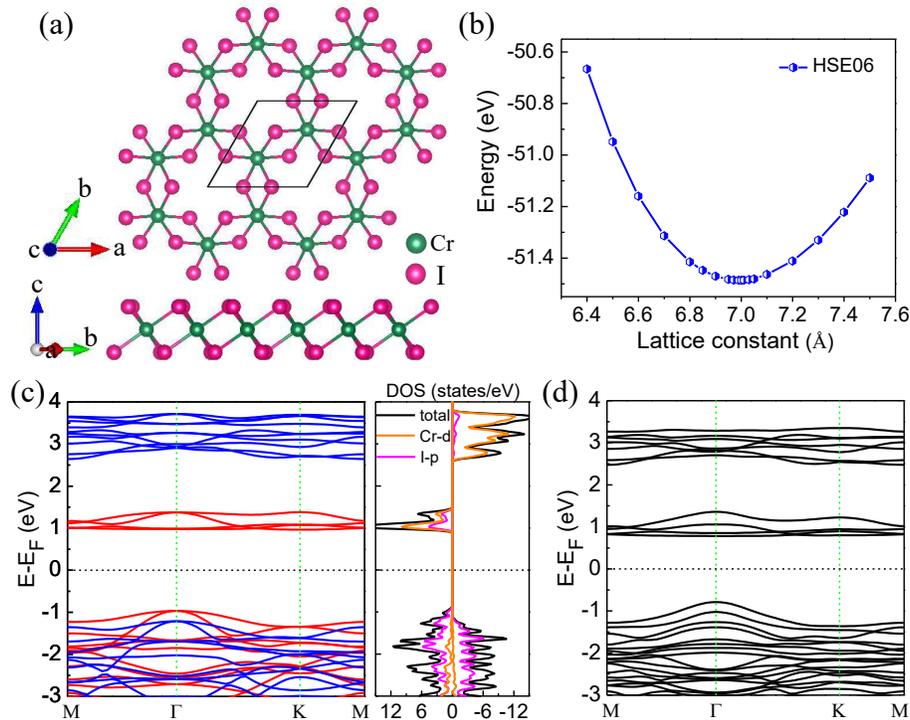}}
\caption{(a) Top and side views of the monolayer CrI$_{3}$. The unit cell is denoted with solid lines. (b) The calculated total energy of the monolayer CrI$_{3}$ as a function of the lattice constant by using HSE06 functional. (c) (Left) The spin-polarized band structure of the monolayer CrI$_{3}$ calculated with HSE06. The red and blue curves denote the spin-up and spin-down bands, respectively. (Right) The densities of states of the monolayer CrI$_{3}$ with the optimized lattice constant obtained from HSE06. (d) The corresponding band structure with the SOC included. Note that the E$_{F}$ in (c) and (d) is set in the middle of the band gap of the monolayer CrI$_{3}$.}
\end{figure*}

The bulk CrI$_{3}$ is a layered compound with van der Waals gaps between adjacent 2D monolayers, which can be cleaved easily and is stable in air \cite{30}. The experimental work shows that CrI$_{3}$ is a FM insulator with a relatively high \emph{T$_{c}$} of 61 K \cite{30}. Recently, the 2D monolayer CrI$_{3}$ has been successfully synthesized in experiments \cite{31} and found to be a FM insulator with an out-of-plane easy axis and a \emph{T$_{c}$} of 45 K. Thus, the FM and insulating behaviors of the bulk CrI$_{3}$ are preserved well even though it is cleaved down to the monolayer limit \cite{31}. In the 2D monolayer CrI$_{3}$, Cr ions are in a honeycomb lattice and coordinated by edge-sharing octahedra with six I ions, as shown in Fig. 1(a). Because the traditional GGA calculation method cannot precisely describe the band gaps of insulators (or semiconductors), we adopt the hybrid functional HSE06 to investigate the geometry and electronic structures of the 2D monolayer FM insulator CrI$_{3}$. The calculations show that the magnetic moment per unit cell of the monolayer CrI$_{3}$ is 6.0 $\mu$$_{B}$. As shown in Fig. 1(b), the obtained relaxed lattice constant in the plane of the monolayer CrI$_{3}$ by HSE06 functional is \emph{a} = 7.00 {\AA}, slightly larger than its bulk lattice constant of \emph{a} = 6.867 {\AA} measured in experiments \cite{30}. The left of Fig. 1(c) displays the calculated spin-polarized band structure of the 2D monolayer CrI$_{3}$ by using HSE06, indicating that the monolayer CrI$_{3}$ is a FM insulator with an indirect band gap of 1.93 eV when the SOC is not included. The top of the valence bands and the bottom of the conduction bands are located at the ¦£ point and K/K$^{\prime}$ points, respectively. As shown on the right of Fig. 1(c) of the densities of states (DOSs) of the system, the top of the valence bands is mainly contributed by the p orbitals of the I atoms. When the SOC is considered, the band gap of the 2D monolayer CrI$_{3}$ become small and is reduced to 1.57 eV. In addition, the monolayer CrI$_{3}$ with SOC included become a direct band gap insulator, with the top of the valence bands and the bottom of the conduction bands both located at the ¦£ point [Fig. 1(d)]. The obtained band gap by GGA+U calculations without and with SOC are 1.19 eV and 0.82 eV, respectively, as displayed in Fig. S1 of the Supplemental Material \cite{38}, which are expectably smaller than the results obtained from the HSE06 functional.

The 2D monolayer FM insulator CrI$_{3}$ has a hexagonal structure in which the magnetic Cr ions form a honeycomb lattice, making the monolayer CrI$_{3}$ match very well with the graphene layer. In our calculations, the heterostructure of Gr/CrI$_{3}$ is built by depositing a $ 5 \times 5$ supercell of graphene on a $\sqrt{3} \times \sqrt{3}$ supercell of the monolayer CrI$_{3}$, as illustrated in Fig. 2(a). The lattice mismatch between the graphene and the CrI$_{3}$ substrate in the constructed heterostructure is only about 1.5 $\%$. When building the heterostructure, we considered three representative interface configurations between graphene and the CrI$_{3}$ substrate: TCr, TI, and Tno. The TCr and TI represent that two C atoms of the top graphene layer lie directly above two Cr and I atoms of the monolayer CrI$_{3}$ in the unit cell of the built heterostructure, respectively, as shown respectively in Fig. 2(a) and Fig. S2(a), while the Tno represents that all of the C atoms of the graphene layer do not lie above any of the Cr or I atoms of the CrI$_{3}$ in the unit cell, as shown in Fig. S2(d). For the TCr configuration, there is also one Cr atom located at the hollow site of graphene (Fig. 2(a)). The obtained binding energies of the heterostructures with the three types of interface configurations are all found to be 36 meV per C atom, indicating that the top layer graphene can be placed on the monolayer CrI$_{3}$ substrate without any site selectivity. The average relaxed distance between graphene and the monolayer CrI$_{3}$ substrate is found to be d$_{0}$ = 3.53 {\AA}, which together with the small binding energy of the heterostructure means the graphene is bonded to the substrate via the weak van der Waals interaction. Therefore, the Gr/CrI$_{3}$ belongs to the so called vdW heterostructures \cite{25,26}, relatively facilely to be fabricated in experiments with current advanced sample synthesis technologies. In the following, the electronic structures and topological properties of the Gr/CrI$_{3}$ with the TCr configuration [as shown in Fig. 2(a)] are systematically investigated, while the calculated results of the heterostructures with the TI and Tno configurations are given in the supporting information, as shown in Fig. S2 and Fig. S3. All the calculations are conducted with GGA+U methods, except as specifically indicated.

\begin{figure*}
\resizebox{17.0cm}{!}{\includegraphics*[58,55][602,358]{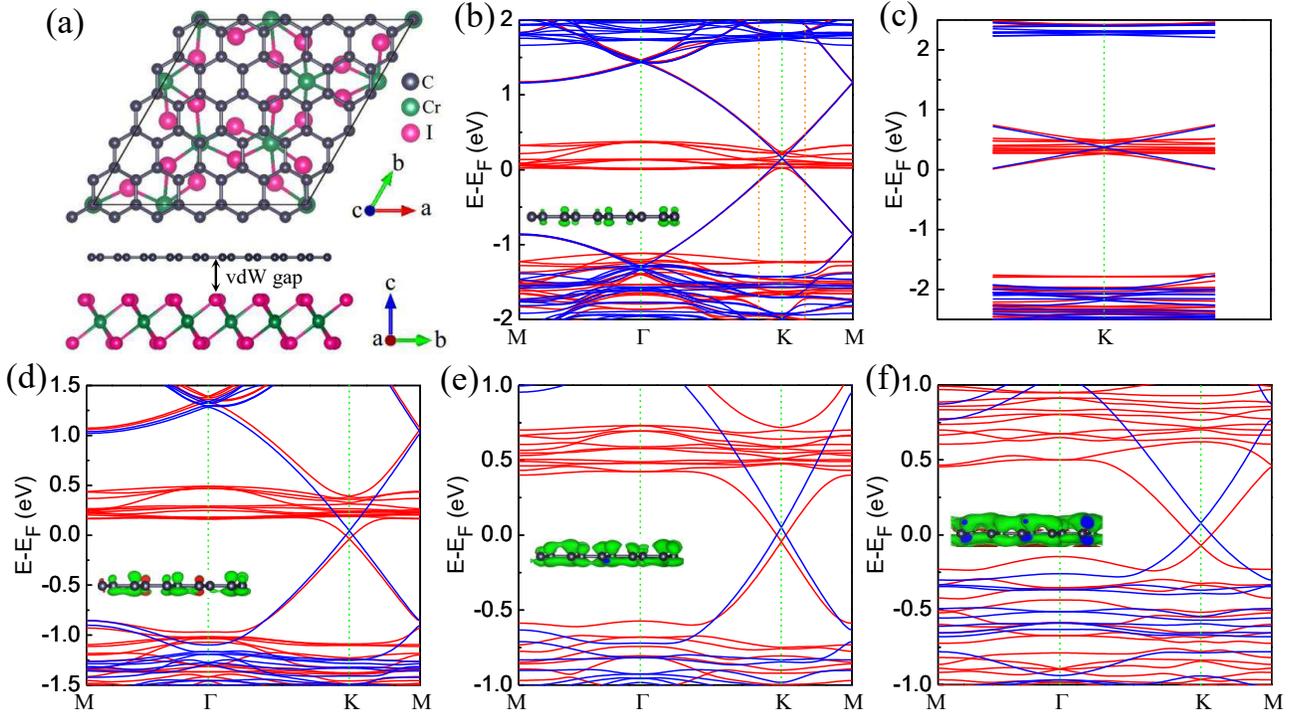}}
\caption{(a) Top and side views of the built Gr/CrI$_{3}$ vdW heterostructure. The distance between graphene and CrI$_{3}$ in the heterostructure is denoted as vdW gap. (b) The spin-polarized band structure of the Gr/CrI$_{3}$ heterostructure with the equilibrium interface distance ($\Delta$d = 0.0 {\AA}) by using GGA+U calculations. Inset: the corresponding spatial distribution of the spin-polarized electron density of the graphene layer in the heterostructure. The red and green colors give the net spin-up and spin-down charge densities, respectively. The isosurface value is set to be 0.0002 e/{\AA}$^{3}$. (c) The calculated band structure around the Dirac K point of the Gr/CrI$_{3}$ in (b) by using the HSE06 functional. (d)-(f) the same as (b), except that the interface distance between graphene and the monolayer CrI$_{3}$ are compressed by $\Delta$d = -0.5 {\AA} (d), $\Delta$d = -0.8 {\AA} (e), and $\Delta$d = -1.1 {\AA} (f), respectively. The red and blue curves denote the spin-up and spin-down bands, respectively.}
\end{figure*}

The calculated spin-polarized band structure of the built vdW heterostructure [Fig. 2(a)] with a reduced vdW gap of d = d$_{0}$ - 0.5 {\AA} (namely $\Delta$d = d - d$_{0}$ = -0.5 {\AA}) is plotted in Fig. 2(d). Amazingly, the Dirac points of graphene in this compressed heterostructure are now located exactly inside the band gap of the monolayer FM insulator substrate CrI$_{3}$, needed to acquire the Chern insulating state. The bands in Fig. 2(d) also demonstrate a very large magnetic exchange splitting (of about 80 meV) around the Dirac points of the graphene induced by the FM CrI$_{3}$ substrate, seen more clearly from the magnified bands plotted in Fig. 4(b). Thus, the slightly reduced vdW gap of the Gr/CrI$_{3}$ heterostructure can lead to a very distinct spin-splitting in the Dirac bands. Because of the weak vdW interaction in the heterostructure interface, the vdW gap should be easily shrunk in experiments by applying an externally vertical compressive pressure. The large magnetic exchange splitting gives rise to the crossing of the spin-up and spin-down bands around the Dirac points [at both K and K$^{\prime}$ points, Fig. 2(d)], essential to achieve the Chern insulating state in the graphene-based systems \cite{7}. If the vdW gap is further compressed, the influence from the substrate increases accordingly, inducing stronger magnetization in the graphene layer, as illustrated in the insets in Fig. 2(e) and Fig. 2(f), and causing larger magnetic exchange splitting in the Dirac bands of the graphene. The exchange splitting becomes 95 and 150 meV when the vdW gap is compressed by 0.8 {\AA} ($\Delta$d = -0.8 {\AA}) and 1.1 {\AA} ($\Delta$d = -1.1 {\AA}), respectively, as seen from the bands in Fig. 2(e) and Fig. 2(f). The very large magnetic exchange splitting in the graphene layer induced by the 2D monolayer FM insulator CrI$_{3}$ in the compressed vdW heterostructure is a very interesting discovery, inferring that the CrI$_{3}$ is a very promising FM insulator substrate for many other 2D target materials.

\subsection{\textbf{Work functions, contact types, and band alignments}}

\begin{figure*}
\resizebox{17.0cm}{!}{\includegraphics*[56,90][600,385]{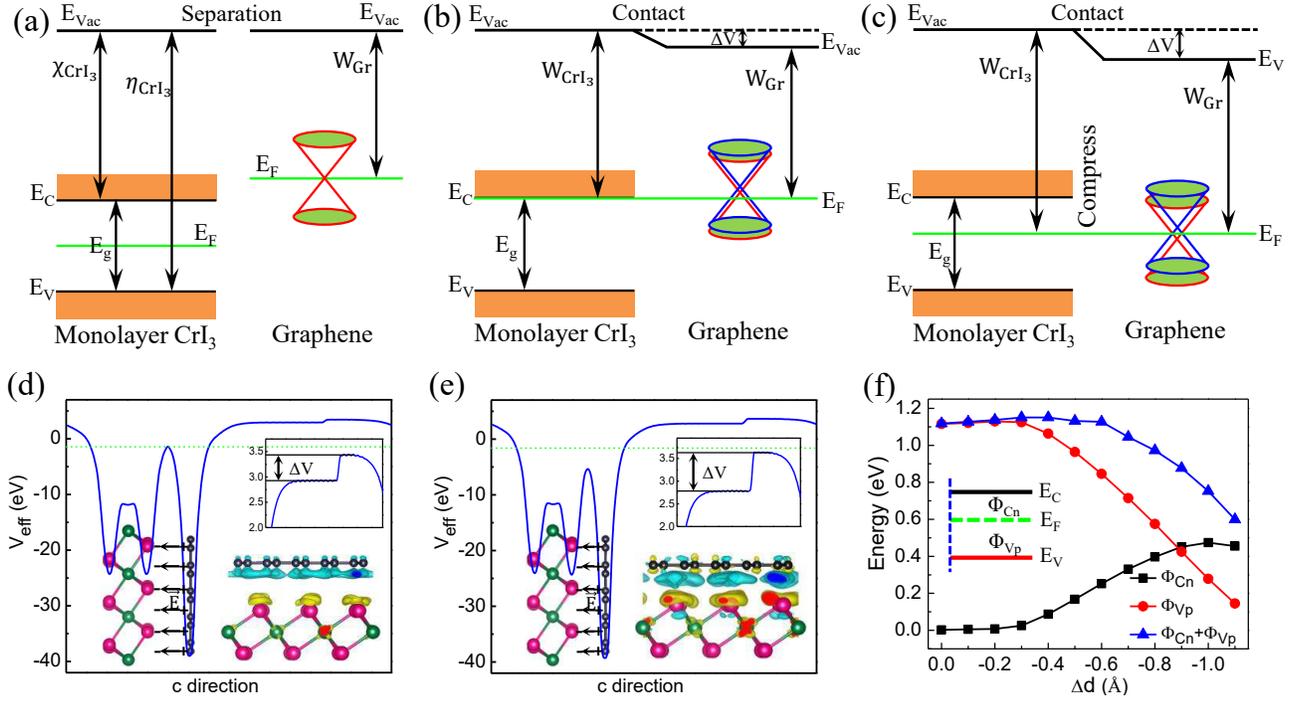}}
\caption{(a)Schematic drawing of the energy levels for the isolated monolayer CrI3 and graphene.W$_{Gr}$ represents the work function of graphene. $\chi$CrI$_{3}$ and $\eta$CrI$_{3}$ represent the electron affinity and ionization potential of the monolayer CrI$_{3}$, respectively. E$_{vac}$ represents the vacuum level. (b) and (c) Schematic drawings of the band alignments for the Gr/CrI$_{3}$ heterostructures with the equilibrium interface distance (b) and with the compressed interface distance (c), respectively. $W_{CrI_{3}}$ represents the work function of the monolayer CrI$_{3}$ in the heterostructure. $\Delta$V represents the potential change generated by the interaction between the monolayer CrI$_{3}$ and graphene.(d) The electrostatic potential (V$_{eff}$) for the Gr/CrI$_{3}$ heterostructure with the equilibrium interface distance ($\Delta$d = 0.0 {\AA}). Insets: (up) the corresponding potential change $\Delta$V, (down) differential charge density of the Gr/CrI$_{3}$ heterostructure. The yellow and cyan colors represent the charge accumulation and depletion, respectively. The direction of the induced internal electric field is also denoted. (e) The same as (d) except for the interface distance is compressed instead, with $\Delta$d = -0.8 {\AA}. The isosurface values in (d) and (e) are set to be 0.0003 e/{\AA}$^{3}$(d) and 0.0008 e/{\AA}$^{3}$(e), respectively. (f) The $\Phi_{Cn}$, $\Phi_{Vp}$, and $\Phi_{Cn}$ + $\Phi_{Vp}$ in the Gr/CrI$_{3}$ heterostructure as a function of $\Delta$d. }
\end{figure*}

\begin{table}
\caption{The first-principles calculations on the work function of the isolated graphene (W$_{Gr}$) and the electron affinity ($\chi_{CrI_{3}}$), ionization potential ($\eta_{CrI_{3}}$), and energy band gap (E$_{g}$) of the isolated monolayer CrI$_{3}$ by using the GGA, GGA+SOC, HSE06, and HSE06+SOC, respectively.}
\begin{ruledtabular}
\begin{tabular}{lccccc}
& Graphene & & CrI$_{3}$ & \\
\hline
& W$_{Gr}$(eV) & $\chi_{CrI_{3}}$(eV) & $\eta_{CrI_{3}}$(eV) & E$_{g}$(eV) \\
\hline
GGA & 4.21 & 4.68 & 5.87 & 1.19 \\
GGA+SOC & 4.21 & 4.70 & 5.52 & 0.82 \\
HSE06 & 4.29 & 4.50 & 6.43 & 1.93 \\
HSE06+SOC & 4.29 & 4.48 & 6.05 & 1.57 \\
\end{tabular}
\end{ruledtabular}
\end{table}

\begin{figure*}
\resizebox{17.0cm}{!}{\includegraphics*[48,70][635,370]{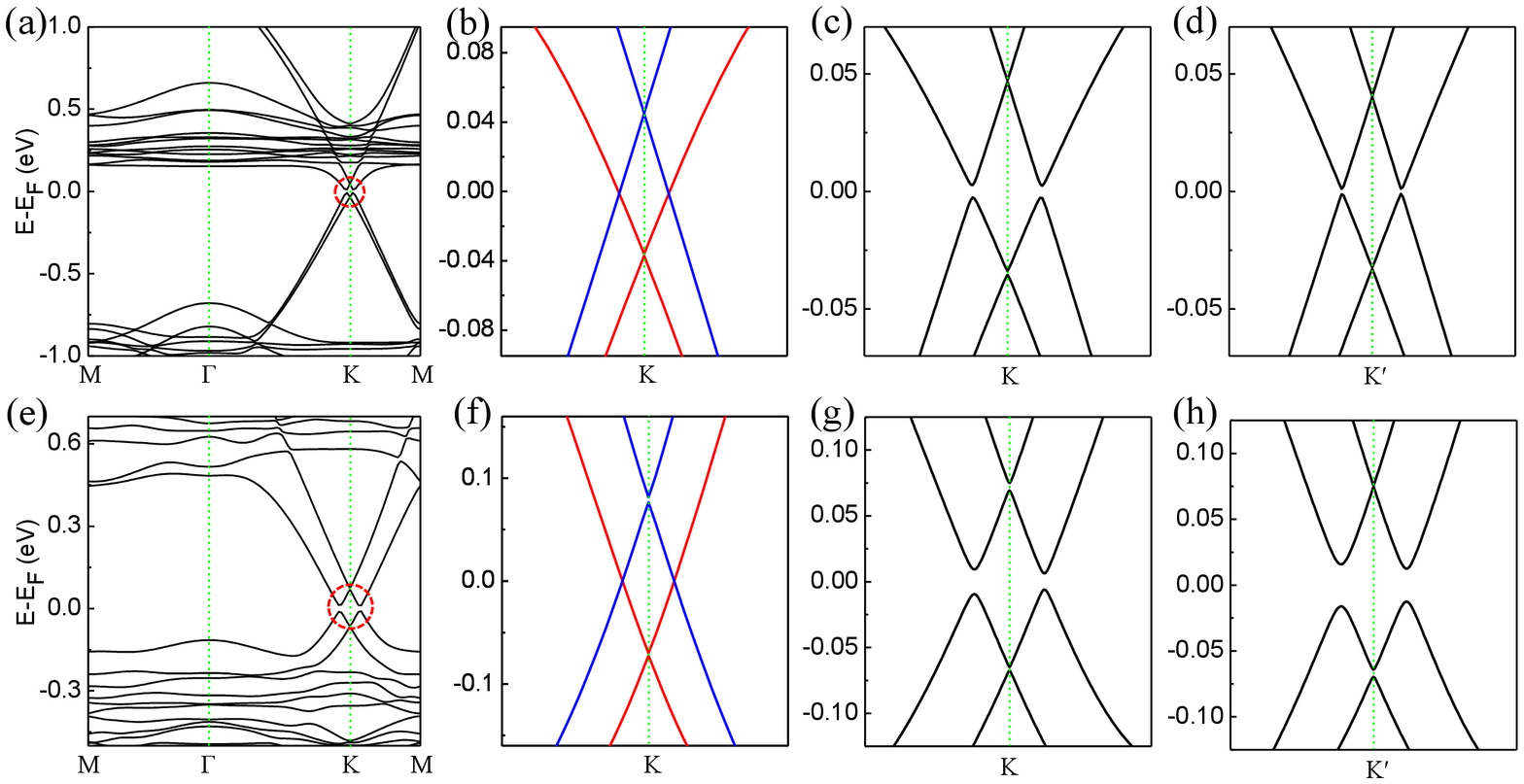}}
\caption{(a) The band structure of the Gr/CrI$_{3}$ heterostructure with $\Delta$d = -0.50 {\AA}. The SOC is involved. (b) The spin-polarized bands around the K point with $\Delta$d = -0.50 {\AA} without the SOC. The red and blue curves denote the spin-up and spin-down bands, respectively. (c) and (d) The corresponding SOC bands around the K and K$^{\prime}$ points, respectively. (c) is also the zooming in of (a) around the Dirac point. (e)-(h) The same as (a)-(d), except for $\Delta$d = -1.10 {\AA} instead.}
\end{figure*}

Figure 3(a) plots a schematic drawing of the energy levels for the isolated monolayer CrI$_{3}$ and graphene, where the work function (the energy difference between the vacuum level (E$_{vac}$) and the E$_{F}$) of graphene (W$_{Gr}$), the electron affinity (the energy difference between E$_{vac}$ and the conduction band minimum (E$_{C}$)), and the ionization potential (the energy difference between E$_{vac}$ and the valence band maximum (E$_{V}$)) of the monolayer CrI$_{3}$ ($\chi$CrI$_{3}$ and $\eta$CrI$_{3}$) are labeled. The first-principles calculations of the W$_{Gr}$, $\chi$CrI$_{3}$ and $\eta$CrI$_{3}$  by using the GGA and HSE06 methods are summarized in Table I. The calculated work function of graphene is 4.29 eV (HSE06+SOC), consistent with previous reports \cite{39}. When these two materials are put together to form a heterostructure, the electrons will flow from the graphene to CrI$_{3}$, due to the electron affinity of the isolated monolayer CrI$_{3}$ is larger than the work function of graphene ($\chi$CrI$_{3}$$>$W$_{Gr}$) \cite{40}. Figure 3(d) presents the calculated electrostatic potential (V$_{eff}$) for Gr/CrI$_{3}$ heterostructure with $\Delta$d = 0.0 {\AA}, from which a potential change $\Delta$V ($\backsim$ 0.5 eV) can be observed. The corresponding differential charge density of the heterostructure is shown in the inset of Fig. 3(d), it is calculated as $\Delta$$\rho$ = $\rho$(Gr/CrI$_{3}$) - $\rho$(graphene) - $\rho$(CrI$_{3}$), where $\rho$(Gr/CrI$_{3}$), $\rho$(graphene), and $\rho$(CrI$_{3}$) are the charge densities of the Gr/CrI$_{3}$ heterostructure, graphene, and monolayer CrI$_{3}$, respectively. In the interface, we can observe charge depletion near graphene and accumulation near the CrI$_{3}$, inducing an internal electric field. The direction of the induced internal electric field in the interface is from graphene to CrI$_{3}$, as denoted in Fig. 3(d). Generally, Schottky or Ohmic contacts are formed at the metal/semiconductor interfaces. According to Schottky-Mott model \cite{41}, an n-type Schottky barrier is defined as the energy difference between the conduction band minimum (E$_{C}$) and the E$_{F}$ ($\Phi_{Cn}$ = E$_{C}$ - E$_{F}$) and a p-type Schottky barrier is defined as the energy difference between the E$_{F}$ and the valence band maximum (E$_{V}$) ($\Phi_{Vp}$ = E$_{F}$ - E$_{V}$). Obviously, the sum of the n-type and p-type Schottky barriers is equal to the band gap (E$_{g}$) of the semiconductor in the heterostructure. For the Gr/CrI$_{3}$ heterostructure with the equilibrium interface distance, an Ohmic contact (n-type) is formed at the Gr/CrI$_{3}$ interface due to W$_{Gr}$ close to $\chi$CrI$_{3}$ (Table I). Based on the above analysis and calculations, the schematic drawing of the band alignment for the Cr/CrI$_{3}$ heterostructure with $\Delta$d = 0.0 {\AA} is plotted in Fig. 3(b).

Figure 3e shows the calculated electrostatic potential (V$_{eff}$) for the compressed Gr/CrI$_{3}$ heterostructure with $\Delta$d = -0.80 {\AA} [also see Fig 3(c)]. The potential change $\Delta$V and the charge redistribution in the heterostructure interface are both enhanced effectively. Figure 3(f) plots the calculated $\Phi$C$_{n}$, $\Phi$V$_{p}$, and $\Phi$C$_{n}$ + $\Phi$V$_{p}$ in the Gr/CrI$_{3}$ heterostructure as a function of $\Delta$d. With the decrease of the vdW gap, the heterostructure E$_{F}$ moves from the E$_{C}$ to E$_{V}$. Thus, the Ohmic contact at the Gr/CrI$_{3}$ interface with the equilibrium distance is tuned first to the n-type Schottky contact and then to the p-type Schottky contact (with $|$$\Delta$d$|$ $>$ 0.8 {\AA}) by reducing the interface distance. In addition, the band gap of the monolayer CrI$_{3}$ in the heterostructure becomes smaller with the decrease of the vdW gap, due to the stronger interaction from the graphene. Therefore, the E$_{F}$ of the heterostructure and the Dirac points of the graphene can be tuned conveniently into the insulating band gap of the CrI$_{3}$ substrate by reducing the interface distance.

\subsection{\textbf{Topological properties}}

\begin{figure*}
\resizebox{17.0cm}{!}{\includegraphics*[72,47][676,397]{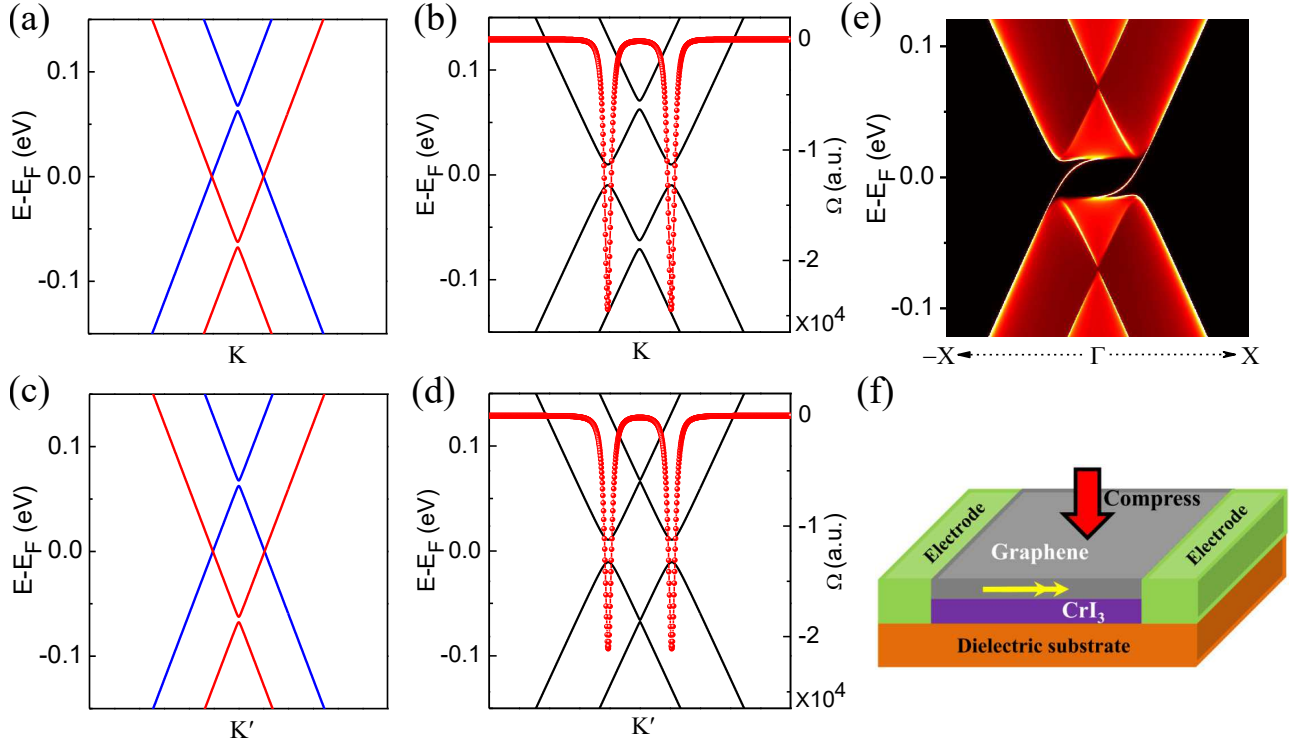}}
\caption{(a) The spin-polarized band structure around the K point obtained from the low-energy continuum model calculations. The red and blue curves denote the spin-up and spin-down bands, respectively. (b) The corresponding band structure (black curves) around the K point with the SOC included. The red dots give the calculated Berry curvatures for the whole valence bands. (c) and (d) The same as (a) and (b) except for around the K$^{\prime}$ point instead. (e) The calculated edge density of states of the semi-infinite armchair-edged graphene system. (f) A schematic diagram depicting the observation of the QAH effect in the vdW heterostructure of Gr/CrI$_{3}$. The vertical red arrow denotes the external compression. The small horizontal yellow arrows indicate the two dissipationless edge current channels owned in the heterostructure.}
\end{figure*}

When the SOC interaction is included in the compressed vdW heterostructure Gr/CrI$_{3}$, an energy gap is opened at the crossing points of the spin-up and spin-down Dirac bands, and the E$_{F}$ is located just inside the SOC-induced gap, as illustrated in Fig. 4(a) and Fig. 4(e). The gap is opened by the enhanced Rashba SOC arose in the top graphene layer due to the existence of the CrI$_{3}$ substrate. Figure 4(c) and Figure 4(d) plot the magnified bands around the K and K$^{\prime}$ points of the compressed heterostructure when the vdW gap is reduced by 0.5 {\AA}. We find that the SOC-induced local band gaps around the K (5.1 meV) and K$^{\prime}$ (2.2 meV) points are not equal, because of the non-equivalent A and B sublattices of graphene, induced by the CrI$_{3}$ substrate. Therefore, a global band gap of about 2.2 meV is opened in this system. If the substrate effect is further enhanced by reducing the vdW gap, the strength of the Rashba SOC will be enlarged accordingly. Figure 4(g) and Figure 4(h) show that if the vdW gap is compressed by 1.1 {\AA}, the enhanced SOC induced local band gaps around the K and K$^{\prime}$ points will be enlarged to 12.2 meV and 25.0 meV, respectively. Therefore, a global SOC-induced band gap of larger than 10 meV can be achieved in the compressed vdW heterostructure, much larger than the nontrivial gap of pure graphene \cite{21,22}. The magnitude of this achieved band gap corresponds to a temperature of higher than 100 K, easily accessible under current experimentally technologies.

We now adopt the low-energy effective model of graphene around the Dirac points to investigate the mechanism of the gap-opening and the topological properties of the above built compressed vdW heterostructure Gr/CrI$_{3}$. The low-energy continuum model of graphene, including the magnetic exchange field, the Rashba SOC, and the staggered AB-sublattice potential, can be written as \cite{12}
\begin{equation*}
H(\mathbf{k})= - \nu_{f}(\eta\sigma_{x}k_{x} + \sigma_{y}k_{y})\textbf{I}_{s} - \emph{M}\textbf{I}_{\sigma}s_{z}
\end{equation*}%
\begin{eqnarray*}
+\frac{\lambda_{R}}{2}(\eta\sigma_{x}s_{y} - \sigma_{y}s_{x}) + \emph{U}\sigma_{z}\textbf{I}_{s},\ \ \ \ \ \ (1)
\end{eqnarray*}%
where $\nu$$_{f}$ = $3t/2$ is the Fermi velocity, $\eta$ = $\pm$ 1 for K and K$^{\prime}$, respectively, and $\sigma$ and \textbf{s} are Pauli matrices that act on the sublattices and the spin degrees of freedom. In Equation 1, the first term represents the nearest-neighboring hopping of graphene with amplitude t = 2.6 eV, the last three terms describe the magnetic exchange field, the Rashba SOC, and staggered sublattice potential, respectively. For the pristine graphene, perfect linear Dirac bands appear at the K and K$^{\prime}$ points, respectively. If the staggered potential \emph{U} is considered, two trivial band gaps are opened around the Dirac points. When a relatively large magnetic exchange field \emph{M} is included, the spin-up and spin-down Dirac bands cross with each other around the K and K$^{\prime}$ points, and the trivial gaps disappear, as displayed in Fig. 5(a) and Fig. 5(c) with \emph{U} = 2.6 meV and \emph{M} = 65 meV. When the Rashba SOC $\lambda$$_{R}$ is further included, two band gaps will be opened at the band crossing points around the Dirac points, as shown in Fig. 5(b) and Fig. 5(d) with \emph{U} = 2.6 meV, \emph{M} = 65 meV, and $\lambda$$_{R}$ = 20.8 meV. Due to the existence of staggered potential, the SOC-induced local band gaps around Dirac points are not equal. Thus, the physical mechanism of the band evolution around the Dirac points of the compressed heterostructure Gr/CrI$_{3}$ can be explained well by this low-energy continuum model of graphene.

The Berry curvatures and Chern number are calculated to identity the topological properties of the above SOC-induced insulating states. The Berry curvature is calculated by \cite{42,43}
\begin{equation*}
\Omega (\mathbf{k})=\sum\limits_{n}{{f_{n}}}{\Omega _{n}}(\mathbf{k}),
\end{equation*}%
\begin{eqnarray*}
\Omega _{n}(\mathbf{k}) &=&-2\mathrm{Im}\sum_{m\neq n}\frac{\hbar ^{2}\left\langle
\psi _{n\mathbf{k}}|v_{x}|\psi _{m\mathbf{k}}\right\rangle \left\langle \psi
_{m\mathbf{k}}|v_{y}|\psi _{n\mathbf{k}}\right\rangle }{%
(E_{m}-E_{n})^{2}},\ \ \ \ \ \ (2)
\end{eqnarray*}%
where the summation is over all of the occupied states, $f_{n}$ is the Fermi-Dirac distribution function, $E_{n}$ is the eigenvalue of the Bloch functions ${\left\vert {{\psi_{n\mathbf{k}}}}%
\right\rangle}$, and $\upsilon _{x(y)}$ are the velocity operators. The Chern number $C$ can be obtained by integrating the Berry curvatures over the first Brillouin Zone (BZ), as $C = \frac{1}{{2\pi }}\sum\limits_{n}{\int_{BZ}{{d^{2}}k}}{\Omega _{n}}$. The calculated $\Omega (\mathbf{k})$ around the K and K$^{\prime}$ points from the low-energy continuum model of graphene are plotted in Fig. 5(b) and Fig. 5(d), respectively. The peaks of the $\Omega (\mathbf{k})$ distribute around the SOC-induced band gaps and they have the same sign near the K and K$^{\prime}$ points. By integrating the Berry curvatures around the K and K$^{\prime}$ points in the first BZ, the nonzero Chern number of $C_{K} = C_{K^{\prime}} = 1$ can be obtained, leading to the total Chern number of $C = C_{K} = C_{K^{\prime}} = 2$. The obtained nonzero integer Chern number indicates that this SOC-induced insulating state is topologically nontrivial and the QAH effect can be realized in this compressed Gr/CrI$_{3}$ vdW heterostructure. Figure 5(e) plots the calculated edge density of states of the semi-infinite armchair-edged graphene system, from which we can observe two distinct chiral edge states in the bulk band gap, consistent with the above obtained total Chern number $C$ = 2. Figure 5(f) gives a schematic diagram depicting the experimental observation of the QAH effect in the built vdW heterostructure of graphene on the monolayer CrI$_{3}$ substrate.
In consideration of the $T_{c}$ of the monolayer CrI$_{3}$ substrate (45 K) and the opened Chern insulator gap in the compressed Gr/CrI$_{3}$ heterostructure (100 K), we can make a statement that the QAH effect can be observed in this heterostructure at a temperature up to 45 K, which is much higher than that in the magnetic topological insulator thin films (30 mK) \cite{17,18,19,20}. The achieved Chern insulating state in the constructed vdW heterostructures also shows strong robustness, as revealed by the band structures of the heterostructures with other two interface configurations, as illustrated in Fig. S2 and Fig. S3. Both of the systems also exhibit the Chern insulating state around the Dirac points.

\section{\textbf{CONCLUSIONS}}

We systematically investigated the electronic structures and topological properties of the vdW heterostructure of graphene on the 2D monolayer FM insulator CrI$_{3}$ from the first-principles calculations and low-energy effective model. The FM CrI$_{3}$ substrate induces a very large magnetic exchange field in the graphene layer, essential to produce the Chern insulating state, which can be enhanced much by applying a vertical external pressure to the heterostructure plane. An Ohmic contact is formed at the heterostructure interface since the work function of graphene is smaller than the electron affinity of the isolated monolayer CrI$_{3}$ substrate. With the decrease of the interface distance, the interface varies from Ohmic contact to n-type Schottky contact (with $|$$\Delta$d$|$ $>$ 0.2 {\AA}) and then to p-type Schottky contact (with $|$$\Delta$d$|$ $>$ 0.8 {\AA}). Very importantly, the Dirac points of graphene are tuned into the insulating gap of the CrI$_{3}$ substrate with a compressive stress to certain extent, which together with the substrate enhanced Rashba SOC in graphene leads to a global sizeable nontrivial band gap ($>$10 meV) opened in the heterostructure system. The calculated Berry curvature, Chern number, and edge states indicate that a Chern insulating state can be achieved in the system, not sensitive to the interface stacking patterns. The experimentally observing temperature of the Chern insulating state in the heterostructure is expected to be up to 45 K, much higher than that in the magnetic topological insulator thin films (30 mK). Our findings may greatly push the experimental observations of the QAH effect in graphene-based systems at high temperatures.

\begin{acknowledgments}
This work was supported by Natural Science Foundation of Jiangsu Province (China) with Grant No. BK20170376, Natural Science Foundation of Higher Education Institutions of Jiangsu Province (China) with Grant Nos. 17KJB140023 and 17KJA140001, and National Natural Science Foundation of China under Grant Nos. 11574051 and 11604134.
\end{acknowledgments}


\end{document}